\documentstyle[prl,aps,amsmath,amssymb]{revtex}
\twocolumn

\begin{document}
%\input epsf
% \draft command makes pacs numbers print
%\twocolumn[\hsize\textwidth\columnwidth\hsize\csname %
%@twocolumnfalse\endcsname
\draft
% repeat the \author\address pair as needed
%\title{{\em Submitted for publication to {\rm Physical Review Letters}} \\
\title{
Determination of Atom-Surface van der Waals Potentials from
  Transmission-Grating Diffraction Intensities} 
\author{R.~E.~Grisenti, W.~Sch\"ollkopf, J.~P.~Toennies}
\address{Max-Planck-Institut f\"ur Str\"omungsforschung, Bunsenstra{\ss}e 10,
  37073 G\"ottingen, Germany} 
\author{G.~C.~Hegerfeldt and T.~K\"ohler}
\address{Institut f\"ur Theoretische Physik, Universit\"at G\"ottingen, 
  Bunsenstra{\ss}e 9, 37073 G\"ottingen, Germany} 
\date{\today} 

\maketitle
%\widetext

\begin{abstract}
  Molecular beams of 
  rare gas atoms and D$_2$ have been diffracted from
  100 nm period
  SiN$_x$ transmission gratings. The relative
  intensities of the diffraction peaks out to the 8th order 
  depend on the diffracting particle and are interpreted in  
  terms of effective slit widths.
  These differences have been analyzed by a new theory
  which accounts for the long-range van der Waals $-C_3/l^3$
  interaction of the particles with the walls of the grating bars. The
  values of the $C_3$ constant
  for two different gratings 
  are in good agreement and the results exhibit the expected linear 
  dependence on the dipole polarizability.
\end{abstract}
%]
%\narrowtext
%
%\begin{multicols}{2}
%\twocolumn
Already in 1932 Lennard-Jones 
\cite{LennardJones}
predicted that the van der Waals interaction of atoms and
molecules with solid surfaces is given by 
\begin{equation}
  \label{vdWpot}
  V=-\frac{C_3}{l^3},\mbox{ }l\gtrsim 10\mbox{ {\AA}}
\end{equation}
where $l$ is the distance from the surface.
This potential plays an important role in understanding virtually all
static (thermodynamical) and dynamical aspects of gas adsorption
phenomena. Despite its importance, very few experimental
determinations of $C_3$ have so far been reported and most of our
present knowledge is based on theoretical estimates
\cite{Kohn}.
The pioneering experiments by Raskin and Kusch on the deflection of Cs
atoms from a conducting metal surface 
\cite{RaskinKusch}
have recently been extended to alkali atoms in high Rydberg states by
measuring the transmission through $8\,\mbox{mm}$ 
long narrow ($2-9\,\mbox{$\mu$m}$)
channels as a function of their principal quantum number $n$
\cite{Meschede}. 
Similar techniques have also been applied to the interaction
of alkali atoms in their ground state 
\cite{Hinds,Aspect}
or in low excited states
\cite{Sandoghdar}.
Although the scattering of many different atoms and molecules
from solid single crystal surfaces has been extensively studied, the
reflection coefficients are relatively insensitive
to the weak long range attractive forces since the collisions are
largely determined by the reflection from the hard repulsive wall
close to the surface
\cite{Hoinkes}.

Here, a new atom optical technique using transmission grating 
diffraction \cite{PritchardEikonal,schoellkopf94} of molecular 
beams is employed. 
%By deflecting the particles passing near the walls of the grating bars,
The van der Waals force causes a change in
the diffraction intensities just as a smaller slit width would.
%Here, diffraction of atoms or molecules from a transmission grating [??]
%is employed as new atom optical technique, which in some respects 
%is related to the original technique of Raskin and Kusch.
%The van der Waals force leads to more diffraction, just as a smaller
%slit width would.
%a 
%deflection of the particles passing
%near the walls of the grating bars, thereby effectively  
%reducing the width of the slits.  
%This modifies the observed intensities of the
%different diffraction orders. 
A newly developed theory makes it possible to interpret
measurements over a
range of different beam energies 
in terms of the potential constant $C_3$. 
For an incident plane wave
the diffraction peak heights depend on the number of illuminated
slits $N$, as $N^2$. With $N=100$ slits the
gain in sensitivity is about four orders of magnitude over
previous experiments.

The measurements were made with
a previously described \cite{schoellkopf94} 
molecular beam diffraction apparatus. 
The beams are produced by a free jet expansion of
the purified gas through a $5\,\mu\mbox{m}$ diameter, 
2 $\mu$m long orifice
from a source chamber at a temperature $T_0$, 
into vacuum
of about $2\times 10^{-4}\, \mbox{mbar}$. 
At $T_0=300\,\mbox{K}$ the source pressure $P_0$
was $140\,\mbox{bar}$ 
for He, Ne, Ar and D$_2$ and $50\,\mbox{bar}$ for Kr. 
At lower source temperature $P_0$ was
reduced to avoid cluster formation.
The atomic 
beams are characterized by narrow
velocity distributions with $\Delta v/v \approx 2.1$ \% 
(He), 5 \% (Ne), 7.6 \% (D$_2$), 7.7 \% (Ar), and 
10 \% (Kr) at $T_0=300$ K, 
where $\Delta v$ and $v$
denote the full half width and the mean value, respectively.
After passing through the 
0.39 mm diameter 
skimmer the beam is collimated by two $10\, \rm
\mu$m wide and $5\,\mbox{mm}$ tall slits $6\, \mbox{cm}$ and $48\,
\mbox{cm}$ downstream from the source before it impinges on the 
silicon nitride (SiN$_x$) transmission grating 
with a grating period of 
$d=100\,\mbox{nm}$ and $5\,\mbox{mm}$ high slits with 
nominal widths of $s_{\rm nom}=50\,\mbox{nm}$
\cite{savasJVSTB95}
placed $2.5\, \mbox{cm}$ behind the second 
collimating slit.  The diffraction
pattern is measured by rotating the electron impact ionization mass
spectrometer detector around an axis parallel to the grating slits.  
A third,
$25\,\mbox{$\mu$m}$ wide slit, $52\,\mbox{cm}$ downstream from the
grating, provides a measured angular resolution of
$70\,\mbox{$\mu$rad}$ (FWHM).

Transmission measurements
with He and Kr atomic beams indicate 
that the grating bars have a truncated
trapezoidal profile (thickness in the beam direction $t$) 
\cite{schoellkopf98,schoellkopf99}
with the narrow face towards the incident beam.
The measured wedge angles $\beta$ and geometrical slit widths 
$s_0$ (see below) are listed in Table \ref{tab:gratings}.

The diffraction measurements are illustrated
in Fig.~\ref{fig:results}
for four inert gases
as a function of the perpendicular wave vector transfer
$\kappa=k\sin\vartheta$, where $\vartheta$ is the diffraction angle. 
The area under the $n$-th order diffraction peak, 
$I_n$, is proportional to the grating
{\em slit function} evaluated
at the diffraction angle of the maximum position, $\vartheta_n$.
For this grating, I, which has equally wide bars and
slits, the zeros of the {\em slit function} coincide 
with the even diffraction orders
\cite{Sommerfeld}, which
are therefore expected to vanish. 
Whereas for He this is almost the case, for the
heavier rare gases, 
an increasing deviation is observed.
For example, the small He intensity ratio 
of the second and third 
order peaks is slightly larger for Ne, 
almost unity in the case of Ar
and, finally, for Kr is greater than one. 
Similar trends are observed for
the ratio of the sixth and fifth order peaks 
and in the ratio of the
most intense zeroth and first orders, 
which increases significantly from about 0.39 
for He to about 0.52 for Kr.

These differences are attributed to the interaction of the 
atoms with the bar walls, Eq.~(\ref{vdWpot}), which 
so far has not been accounted for in the theory of 
atom/molecule diffraction.
For a plane wave
$e^{ikz}$ 
incident on a 
transmission grating  
with perfectly reflecting
grating bars and
with an additional (attractive) potential at the bar sides, 
the diffracted
wave function is, for large $r$,
\begin{equation}
  \psi({\bf r})\underset{r\to\infty}{\longrightarrow} 
  f(\vartheta)\frac{e^{i(kr-\pi/4)}}{\sqrt{r}},
\end{equation}
where $r^2=x^2+z^2$
is in the scattering plane normal to the height of the slits. 
The scattering amplitude $f(\vartheta)$  
is determined by the grating
transmission function $\psi(x,0)$, i.~e.~by 
the wave
function at the far side slit boundaries ($z=0$),
which depends on the attractive potential. 
Huygens' principle \cite{Sommerfeld}
yields
\begin{equation}
  \label{ampgen}
  f(\vartheta)=\frac{\cos\vartheta}{\sqrt{\lambda}}
  \int_{\rm slits}dx\psi(x,0)e^{-ikx\sin\vartheta}.
\end{equation}
If the slit and the bar widths are much larger
than the de Broglie wave length $\lambda$, the intensity
$I(\vartheta)=|f(\vartheta)|^2$ can be written as a product
\begin{equation}
  \label{diffrint}
  I(\vartheta)=
  \left(
    \frac{\sin\left(\frac{1}{2}Nkd\sin\vartheta\right)}
    {\sin\left(\frac{1}{2}kd\sin\vartheta\right)}
  \right)^2
  \left|f_{\rm slit}(\vartheta)\right|^2,
\end{equation}
where $N$ denotes the number of 
slits and 
$|f_{\rm slit}|^2$ is the
{\em slit function}. Thus, the atomic diffraction pattern consists
of principal maxima at the diffraction angles
$\sin\vartheta_n=n\lambda/d$,
$n=0,\pm 1,\pm 2,\ldots$
while $\left|f_{\rm slit}(\vartheta)\right|^2$ plays the role of an
envelope function.
Eq.~(\ref{ampgen}) gives, after a change of variable from $x$ to
a variable with the origin at the edge of a slit,
$\zeta\equiv s_0/2-x$, 
\begin{equation}
  \label{slitamp1}
  f_{\rm slit}(\vartheta)=\frac{\cos\vartheta}{\sqrt{\lambda}}
  2\int_0^{\frac{s_0}{2}}
  d\zeta\cos
  \left[
    \kappa
    \left(
      \frac{s_0}{2}-\zeta
    \right)
  \right]
  \tau(\zeta),
\end{equation}
where 
$\tau(\zeta)=\psi(s_0/2-\zeta,0)$, $0\leq\zeta\leq s_0/2$, is
the single-slit transmission function. 

It is instructive to first deduce
the general structural
form of $f_{\rm slit}(\vartheta)$.
Since the grating bars reflect those atoms 
which touch the bar walls, the wave
function in the slit vanishes at the walls, i.~e.~$\tau(0)=0$. 
Taking this into account and after a partial
integration Eq.~(\ref{slitamp1}) becomes
\begin{equation}
  \label{slitamp2}
  f_{\rm slit}(\vartheta)=\frac{\cos\vartheta}{\sqrt{\lambda}}
  \tau\left(\frac{s_0}{2}\right)
  \frac{e^{i\kappa\frac{s_0}{2}}\Phi(-\kappa)
    -e^{-i\kappa\frac{s_0}{2}}\Phi(\kappa)}{i\kappa},
\end{equation}
where 
\begin{equation}
  \Phi(\pm\kappa)\equiv
  \int_0^{\frac{s_0}{2}}
  d\zeta
  e^{\pm i\kappa\zeta}
  \frac{\tau'(\zeta)}{\tau\left(\frac{s_0}{2}\right)},
\end{equation}
with $\Phi(0)=1$.
The logarithm of $\Phi$ can be expanded as
%in a power series,
\begin{equation}
  \label{cumexp}
  \log\Phi(\pm\kappa)=\sum_{n=1}^\infty
  \frac{(\pm i\kappa)^n}{n!}R_n,
\end{equation}
where the complex $R_n$ are known as cumulants
\cite{Abramowitz},
\begin{eqnarray}
  \label{expacoeff}
  R_1 &=& \int_0^{\frac{s_0}{2}}
  d\zeta\zeta\frac{\tau'(\zeta)}{\tau\left(\frac{s_0}{2}\right)}
  %\\
  %\nonumber
  = \frac{s_0}{2}-\int_0^{\frac{s_0}{2}}
  d\zeta
  \frac{\tau(\zeta)}{\tau\left(\frac{s_0}{2}\right)},
%  \\
%  \nonumber
%  R_2 &=&
%  \int_0^{\frac{s_0}{2}}
%  d\zeta\zeta^2\frac{\tau'(\zeta)}{\tau\left(\frac{s_0}{2}\right)}
%  -R_1^2\\
%  \nonumber
%  &=&
%  \frac{s_0^2}{4}-R_1^2
%  -2\int_0^{\frac{s_0}{2}}
%  d\zeta\zeta\frac{\tau(\zeta)}{\tau\left(\frac{s_0}{2}\right)},
\end{eqnarray}
etc..
For the small wave-vector transfer $\kappa$ of interest here, 
only the first two terms are needed in the series
Eq.~(\ref{cumexp}). 
The single-slit amplitude
Eq.~(\ref{slitamp2}) then becomes
\begin{equation}
  \label{slitamp3}
  f_{\rm slit}(\vartheta)=2\frac{\cos\vartheta}{\sqrt{\lambda}}
  \tau\left(\frac{s_0}{2}\right)
  e^{-\frac{\kappa^2}{2}R_2}
  \frac{\sin
    \left[
      \kappa
      \left(
        \frac{s_0}{2}-R_1
      \right)
    \right]}{\kappa}.
\end{equation}

For a comparison with experiment the surface roughness of the grating
bars must be accounted for.  
In a first approximation roughness has been included by rigid shifts
of the individual
bar sides (see also Ref.~\cite{Toigo}), 
which are randomly Gaussian distributed.
In the case of a weak surface potential, this
results in an additional Debye-Waller like damping factor 
$\exp(-k^2\sigma_0^2\sin^2\vartheta_n)$ in the intensity ratio  
of the principal maxima, $I_n/I_0$,
where $\sigma_0^2$ is the variance of the geometrical slit width
\cite{schoellkopf99}. Taking this into account, 
Eq.~(\ref{diffrint}) with
Eq.~(\ref{slitamp3}) yields
\begin{equation}
  \label{cumresult}
  \frac{I_n}{I_0}=\frac{e^{-\left(\frac{2\pi n\sigma}{d}\right)^2}}
  {\left(\frac{\pi n\sqrt{s_{\rm eff}^2+\delta^2}}{d}\right)^2}
  \left[
    \sin^2\left(\frac{\pi ns_{\rm eff}}{d}\right)
    +\sinh^2\left(\frac{\pi n\delta}{d}\right)
  \right],
\end{equation}
where $\sigma^2\equiv\sigma_0^2+{\rm Re}(R_2)$,
$s_{\rm eff}\equiv s_0-2{\rm Re}(R_1)$ and 
$\delta\equiv 2{\rm Im}(R_1)$. 
The first term in the brackets
of Eq.~(\ref{cumresult}) 
leads to a Kirchhoff-like slit function 
(see e.~g.~Ref.~\cite{schoellkopf99})
with a Debye-Waller term and an effective reduced slit
width $s_{\rm eff}$, 
while the second term 
suppresses the zeros of the Kirchhoff pattern,
as can be seen in the insets of Fig.~\ref{fig:results}.
 
The effective variance $\sigma^2$ as well as $s_{\rm eff}$ and
$\delta$ in Eq.~(\ref{cumresult}) 
can be calculated for the potential Eq.~(\ref{vdWpot}).
The standard eikonal approximation
\cite{Landau,PritchardEikonal}
is used to determine the  
grating transmission function, given by
$\psi(x,0)=e^{i\varphi(x)}$
in the slits and zero elsewhere. The phase shift reads
\begin{equation}
  \label{eikonalphase}
  \varphi(x)=-\frac{1}{\hbar v}\int dzV(x,z),
\end{equation}
where $v=\hbar k/m$ is the particle velocity.
Taking the trapezoidal bar profile into account, after some
algebra the single-slit transmission function becomes
\begin{equation}
  \label{transf}
  \tau(\zeta)=\exp
  \left[
    i\frac{t\cos\beta}{\hbar v}\frac{C_3}{\zeta^3}
    \frac{1+\frac{t}{2\zeta}\tan\beta}
    {
      \left(
        1+\frac{t}{\zeta}\tan\beta
      \right)^2}
  \right].
\end{equation}
An analysis of Eqs.~(\ref{transf}) and (\ref{expacoeff})
reveals that ${\rm Re}(R_1)$ and hence $s_{\rm eff}$
is especially sensitive to the
potential. 
 
The effective slit width $s_{\rm eff}$ as well 
as $\delta$ and $\sigma$ were determined 
from the experiment by
fitting the relative experimental diffraction intensities 
$I_n/I_1$ as
depicted in the insets of Fig.~\ref{fig:results}
to the corresponding ratios determined from Eq.~(\ref{cumresult}).
These ratios and not $I_n/I_0$ are compared with theory
since small concentrations of clusters in the beams can falsify the 
$I_0$ intensities.
The effective slit widths are
plotted versus the particle velocity 
in Fig.~\ref{fig:seff} (points) for two different gratings.
The difference between the effective slit widths for $T_0=300\,\mbox{K}$ 
beams and the geometrical slit width $s_0$ increases
from $1\,\mbox{nm}$ (He) to more than $6\,\mbox{nm}$ for Kr as 
expected from the 
increasing interaction strength of the van der Waals potential.
With increasing  $C_3$ the slope of the curves also increases.  
The solid lines in Fig.~\ref{fig:seff} represent least
squares fits of the theoretical
expression $s_{\rm eff}=s_0-2{\rm Re}(R_1)$, with $R_1$ given by
Eqs.~(\ref{expacoeff}) and (\ref{transf}), 
to the experimentally determined
effective slit widths, which allow
for the determination of $C_3$ and $s_0$.   
Since He has the smallest polarizability 
and measurements over the largest 
range of velocities were possible
they were used to determine the values of $s_0$ in 
Table \ref{tab:gratings}
for each of the gratings.
Identical values for $s_0$ were obtained from  
D$_2$ measurements.
This value of $s_0$ was then 
used for Ne, Ar and Kr, with $C_3$ 
the only remaining
fit parameter, and hence for these systems 
measurements at various velocities are not
necessary.

The $C_3$ parameters are 
plotted versus the static electric dipole 
polarizabilities $\alpha$ in Fig.~\ref{fig:C3pol}.
The error bars were
determined by assuming a realistic
uncertainty in the bar geometry by
varying $\beta$ by $\pm2^\circ$
in Eq.~(13). This
uncertainty seems to be the only systematic
source of error in the present $C_3$
determination and leads to errors of about 20 \%.
Figure~\ref{fig:seff} indicates that the influence of the surface
potential is restricted to distances much smaller than the slit width and
therefore, by Ref.~\cite{Zhou}, corrections due to the finite bar 
width should be
negligible.
%Corrections of the potential in Eq.~(\ref{vdWpot})
%due to the fact that the bars are not infinitely wide
%should be negligible \cite{Zhou},
%since only those particles are essentially deflected,
%which pass near the sides of the grating bars at distances much
%smaller than the width of the bars.

Within the error bars
the data from both gratings fall on a
straight line in 
agreement with Hoinkes' empirical rule \cite{Hoinkes}. 
Accordingly the slope provides information
on the optical dielectric constant 
of the grating material.
An approximation to the theoretical expression for 
$C_3$ \cite{Cole} predicts that 
D$_2$ should in fact 
have a slightly smaller ratio of 
$C_3/\alpha$ than the rare gas atoms,
while among them Ne is expected to have the largest ratio.
It is satisfying to see that the small deviations 
from the straight line in Fig.~\ref{fig:C3pol}
agree with this expected trend. 

The big advantage of the present method is its large 
sensitivity as can be seen from Fig.~\ref{fig:seff}
and its universality. In principle all atoms and molecules 
are accessible for study. The only restrictions will be to 
produce gratings of different solids and molecular 
beams with sufficiently narrow velocity distributions
and to reduce the corresponding 
background in the mass spectrometer detector
to assure an adequate signal to noise ratio. 
The present work also
allows for a quantitative understanding 
of diffraction intensities in atom optics and atom
interferometry experiments using transmission structures as optical
elements.

We are greatly indebted to Tim Savas 
and Henry I. Smith of MIT for providing
the transmission gratings to us.  Further, we thank Dick Manson
and G. Schmahl for fruitful
discussions.

%\end{multicols}
%\onecolumn
%%%%%%%%%%%%%%%%%%%%%%%%%%%%%%%%%%%%%%%%%%%%%%%%%%%%%%%%%%%%%%%%%%%%%
%%  FIGURE 1:
%%%%%%%%%%%%%%%%%%%%%%%%%%%%%%%%%%%%%%%%%%%%%%%%%%%%%%%%%%%%%%%%%%%%%
\begin{figure}
%\begin{center}
%\epsfysize=21cm
%\leavevmode
%\vspace{2cm}
%\epsfbox{conf1.eps}
%\end{center}
\caption{Diffraction patterns measured with Grating I
  for He, Ne, Ar, and
  Kr at the
  same beam energy ($T_0 = 300\,\mbox{K}$).
  The insets contain a  
  comparison between least-squares fits of $I_n/I_1$
  determined from
  Eq.~(\ref{cumresult}) 
  with continuous values of $n$ (solid lines)
  and Kirchhoff theory (dashed lines) to 
  measured diffraction
  intensity ratios (points).
}
\label{fig:results}
\end{figure}
%%%%%%%%%%%%%%%%%%%%%%%%%%%%%%%%%%%%%%%%%%%%%%%%%%%%%%%%%%%%%%%%%%%%%%%
%%  FIGURE 2:
%%%%%%%%%%%%%%%%%%%%%%%%%%%%%%%%%%%%%%%%%%%%%%%%%%%%%%%%%%%%%%%%%%%%%%%
\begin{figure}
%\begin{center}
%\epsfysize=21cm
%\leavevmode
%\vspace{1cm}
%\epsfbox{}
%\end{center}
\caption{Effective slit widths plotted as a function of the particle
velocity for He, Ne, D$_2$, Ar, and Kr beams.
The solid lines are theoretical curves determined from 
Eqs.~(\ref{expacoeff}) and (\ref{transf}) 
with the $C_3$ parameters in Fig.~\ref{fig:C3pol}.
Data points indicate fits of $I_n/I_1$
determined from Eq.~(\ref{cumresult})
to experimental intensity ratios obtained from diffraction
measurements with two gratings. 
\label{fig:seff}}
\end{figure}
%%%%%%%%%%%%%%%%%%%%%%%%%%%%%%%%%%%%%%%%%%%%%%%%%%%%%%%%%%%%%%%%%%%%%%%
%%  FIGURE 3:
%%%%%%%%%%%%%%%%%%%%%%%%%%%%%%%%%%%%%%%%%%%%%%%%%%%%%%%%%%%%%%%%%%%%%%%
\begin{figure}
%\begin{center}
%\epsfysize=21cm
%\leavevmode
%\vspace{1cm}
%\epsfbox{conf1.eps}
%\end{center}
\caption{Measured $C_3$ values of silicon nitride
  (SiN$_x$) obtained in this work  
  plotted as a function of the static electric
  dipole polarizability 
  of the respective atom, $\alpha$ 
  (see Ref.~\protect\cite{Hoinkes}). 
  The solid line is a
  linear fit of the data.
\label{fig:C3pol}}
\end{figure}
%%%%%%%%%%%%%%%%%%%%%%%%%%%%%%%%%%%%%%%%%%%%%%%%%%%%%%%%%%%%%%%%%%%%%%%
%%  END  FIGURES
%%%%%%%%%%%%%%%%%%%%%%%%%%%%%%%%%%%%%%%%%%%%%%%%%%%%%%%%%%%%%%%%%%%%%%%
%%%%%%%%%%%%%%%%%%%%%%%%%%%%%%%%%%%%%%%%%%%%%%%%%%%%%%%%%%%%%%%%%%%%%
%%  TABLE 1:
%%%%%%%%%%%%%%%%%%%%%%%%%%%%%%%%%%%%%%%%%%%%%%%%%%%%%%%%%%%%%%%%%%%%%
\begin{table}
\caption{Geometrical properties of the three gratings.
\label{tab:gratings}}
\begin{center}
\begin{tabular}{|c|c|c|}
grating & $\beta$ [$^\circ$] & $s_0$ [nm]\\
\hline
I   & 7.5$\pm 2$  & 50\\
II  & 8.7$\pm 2$    & 67.5$\pm$0.1\\
III & 12.7$\pm 2$  & 71.2$\pm$0.1\\
\hline
\end{tabular}
\end{center}
\end{table}
%%%%%%%%%%%%%%%%%%%%%%%%%%%%%%%%%%%%%%%%%%%%%%%%%%%%%%%%%%%%%%%%%%%%%%%
%%  END  TABLES
%%%%%%%%%%%%%%%%%%%%%%%%%%%%%%%%%%%%%%%%%%%%%%%%%%%%%%%%%%%%%%%%%%%%%%%
%\end{multicols}

\begin{thebibliography}{99}
\bibitem{LennardJones}
  J.~E.~Lennard-Jones, 
  Trans.~Faraday Soc.~{\bf 28}, 334 (1932).
\bibitem{Kohn}
%  See e.~g.~
%  W.~Kohn, in: {\em Interaction of Atoms and Molecules with Solid
%    Surfaces}, Eds.~V.~Bortolani, N.~H.~March and M.~P.~Tosi, (Plenum,
%  New York, 1990).
  See e.~g.~G.~Vidali, G.~Ihm, H.~Y.~Kim, and M.~W.~Cole, 
  Surf.~Sci.~Reports {\bf 12}, 133 (1991).
\bibitem{RaskinKusch}
  D.~Raskin and P.~Kusch, Phys.~Rev.~{\bf 179}, 712 (1969).
  See also A.~Shih and V.~A.~Parsegian, Phys.~Rev.~A {\bf 12},
  835 (1975).
\bibitem{Meschede}
  A.~Anderson, S.~Haroche, E.~A.~Hinds, W.~Jhe, and D.~Meschede,
  Phys.~Rev.~A {\bf 37}, 3594 (1988).
\bibitem{Hinds}
  C.~I.~Sukenik, M.~G.~Boshier, D.~Cho, V.~Sandoghdar, and 
  E.~A.~Hinds, Phys.~Rev.~Lett.~{\bf 70}, 560 (1993).
\bibitem{Aspect}
  A.~Landragin, J.~Y.~Courtois, G.~Labeyrie,
  N.~Vansteen\-kiste, C.~I.~Westbrook, and A.~Aspect,
  Phys.~Rev.~Lett. {\bf 77}, 1464 (1996).
\bibitem{Sandoghdar}
  V.~Sandoghdar, C.~I.~Sukenik, E.~A.~Hinds, and S.~Haroche,
  Phys.~Rev.~Lett.~{\bf 68}, 3432 (1992).
%\bibitem{Aspect}
%  A.~Landragin, J.~Y.~Courtois, G.~Labeyrie,
%  N.~Vansteen\-kiste, C.~I.~Westbrook, and A.~Aspect,
%  Phys.~Rev.~Lett. {\bf 77}, 1464 (1996).
\bibitem{Hoinkes}
  H.~Hoinkes, Rev.~Mod.~Phys.~{\bf 52}, 933 (1980).
\bibitem{PritchardEikonal}
  C.~R.~Ekstrom, D.~W.~Keith, and D.~E.~Pritchard,
  Appl.~Phys.~B {\bf 54}, 369 (1992).
\bibitem{schoellkopf94} 
  W.~Sch\"ollkopf and J.~P.~Toennies,
  Science {\bf 266}, 1345 (1994).
%\bibitem{schoellkopf98}
%  W.~Sch\"ollkopf, J.~P.~Toennies, T.~A.~Savas, and H.~I.~Smith,
%  J.~Chem.~Phys.~{\bf 109}, 9252 (1998).
\bibitem{savasJVSTB95} 
  T.~A.~Savas, S.~N.~Shah, M.~L.~Schattenburg,
  J.~M.~Carter, and H.~I.~Smith,
  J.~Vac.~Sci.~Technol.~B {\bf 13}, 2732 (1995).
\bibitem{schoellkopf98}
  W.~Sch\"ollkopf, J.~P.~Toennies, T.~A.~Savas, and H.~I.~Smith,
  J.~Chem.~Phys.~{\bf 109}, 9252 (1998).
\bibitem{schoellkopf99}
  R.~E.~Grisenti, W.~Sch\"ollkopf, J.~P.~Toennies, J.~R.~Manson,
  T.~A.~Savas, and H.~I.~Smith, submitted to Phys.~Rev.~A.
\bibitem{Sommerfeld}
  A.~Sommerfeld, {\em Optics}, (Academic Press,
  New York, 1950).
\bibitem{Abramowitz}
  M.~Abramowitz and I.~A.~Stegun, {\em Handbook of Mathematical
    Functions}, (Dover Publications, New York, 1972).
\bibitem{Toigo}
  A.~M.~Marvin and F.~Toigo, Phys.~Rev.~A {\bf 25}, 782 (1982).
%\bibitem{Sommerfeld}
%  A.~Sommerfeld, {\em Optics}, (Academic Press,
%  New York, 1950).
\bibitem{Landau}
  L.~D.~Landau, E.~M.~Lifschitz, {\em Quantum Mechanics
    (Nonrelativistic theory)}, Vol.~3, (Pergamon, New York, 1977),
  pp.~538.
%\bibitem{PritchardEikonal}
%  C.~R.~Ekstrom, D.~W.~Keith, and D.~E.~Pritchard,
%  Appl.~Phys.~B {\bf 54}, 369 (1992).
%\bibitem{Abramowitz}
%  M.~Abramowitz and I.~A.~Stegun, {\em Handbook of Mathematical
%    Functions}, (Dover Publications, New York, 1972).
%\bibitem{schoellkopf99}
%  R.~E.~Grisenti, W.~Sch\"ollkopf, J.~P.~Toennies, J.~R.~Manson,
%  T.~A.~Savas, and H.~I.~Smith, submitted to Phys.~Rev.~A.
%\bibitem{Toigo}
%  A.~M.~Marvin and F.~Toigo, Phys.~Rev.~A {\bf 25}, 782 (1982).
%\bibitem{savasJVSTB95} 
%  T.~A.~Savas, S.~N.~Shah, M.~L.~Schattenburg,
%  J.~M.~Carter, and H.~I.~Smith,
%  J.~Vac.~Sci.~Technol.~B {\bf 13}, 2732 (1995).
\bibitem{Zhou}
  F.~Zhou and L.~Spruch, Phys.~Rev.~A {\bf 52}, 297 (1995).
\bibitem{Cole}
  G.~Vidali and M.~W.~Cole, Surf.~Sci.~{\bf 110}, 10 (1981).
%\bibitem{Zhou}
%  F.~Zhou and L.~Spruch, Phys.~Rev.~A {\bf 52}, 297 (1995).
\end{thebibliography}
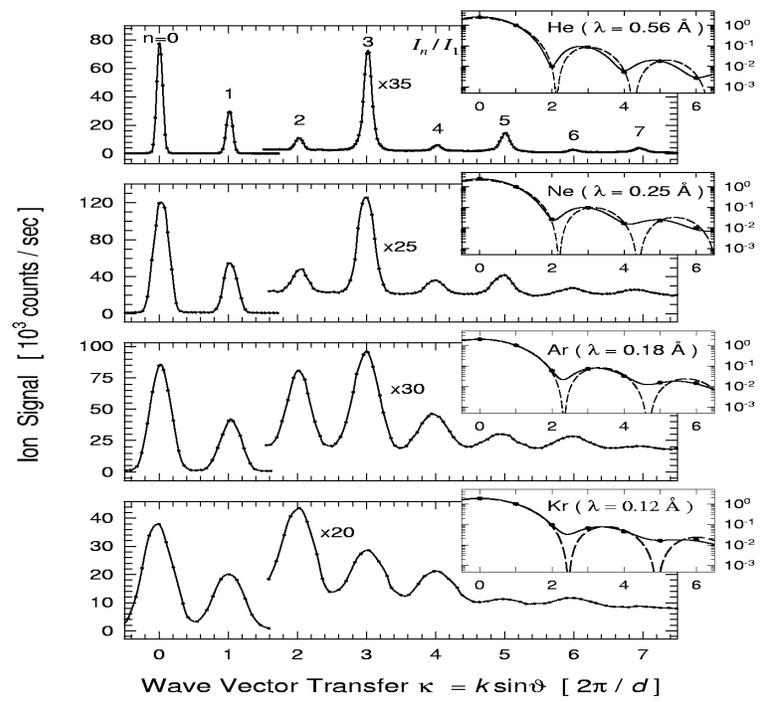
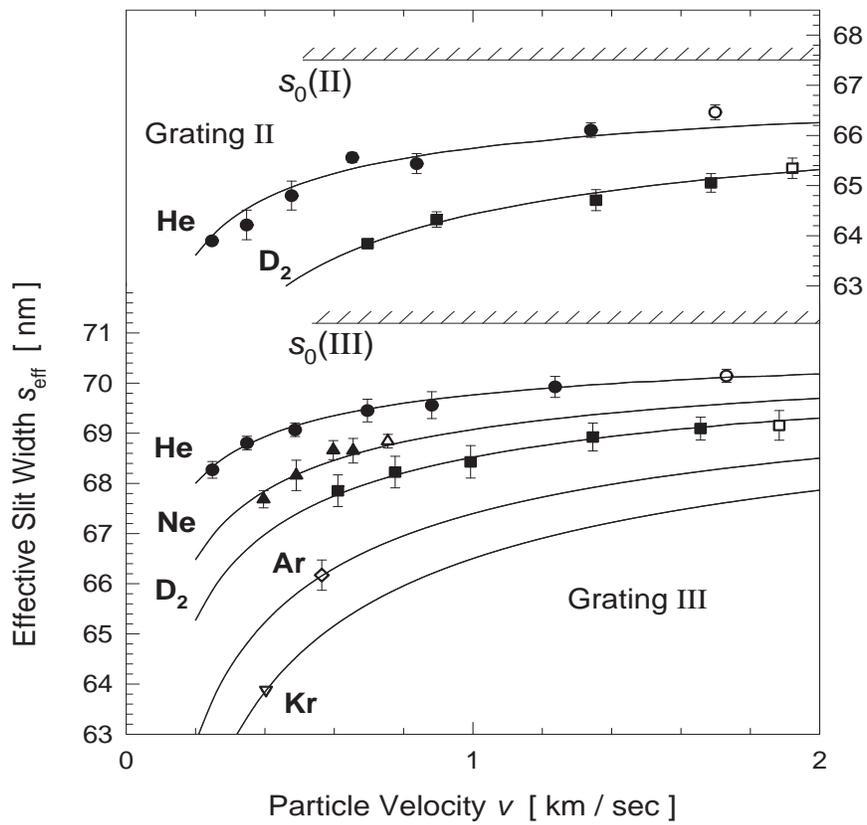
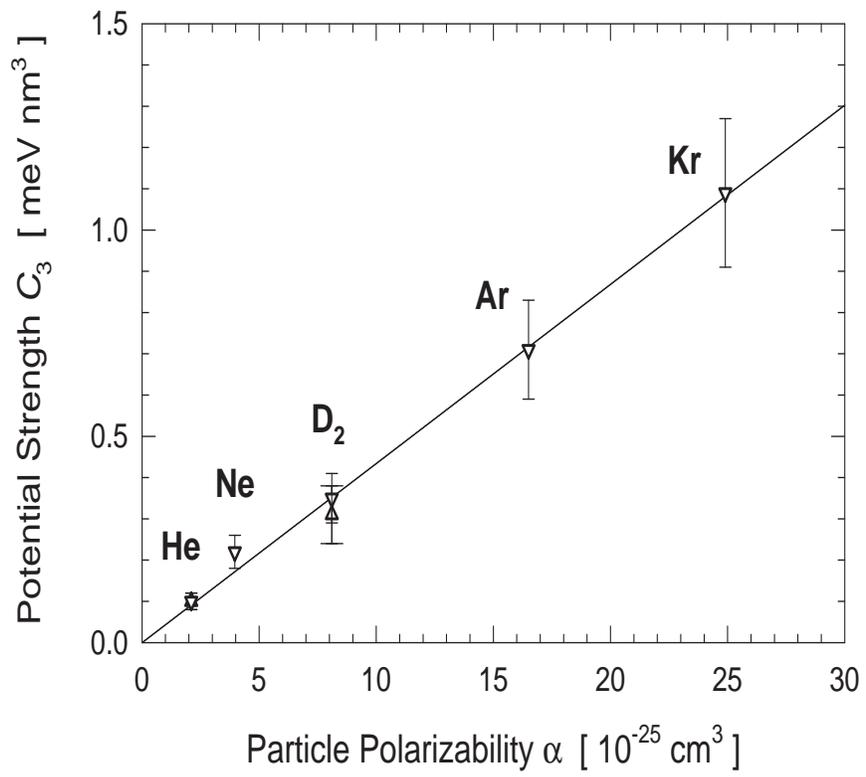
\end{document}